\documentclass[fleqn,twoside]{article}
\usepackage{espcrc2}
\usepackage{amssymb}
\usepackage{graphicx}

\usepackage[figuresright]{rotating}

\newcommand{\AmS}{{\protect\the\textfont2
  A\kern-.1667em\lower.5ex\hbox{M}\kern-.125emS}}

\title{Density profiles of small Dirac operator eigenvalues 
for two color QCD at nonzero chemical potential compared to matrix models
}

\author{Gernot Akemann\address{%
Service de Physique Th\'eorique, CEA/DSM/SPhT Saclay, 
Unit\'e associ\'ee CNRS/SPM/URA 2306,\\
\ \,F-91191 Gif-sur-Yvette Cedex, France}\address{%
Department of Mathematical Sciences, Brunel University West London, Uxbridge, UB8 3PH, UK},
        Elmar Bittner\address{%
Institut f\"ur Theoretische Physik, Universit\"at Leipzig, Augustusplatz 10/11,
 D-04109 Leipzig, Germany},
        Maria-Paola Lombardo\address{%
INFN-Laboratori Nazionali di Frascati,
I-00044 Frascati, Italy},
 Harald Markum\address{Atominstitut, Technische Universit\"at Wien, A-1040
 Wien, Austria},
       Rainer Pullirsch$^{\rm e}$}

\begin{document}

\begin{abstract}
We investigate the eigenvalue spectrum of the staggered Dirac matrix in
two color QCD at finite chemical potential. 
The profiles of complex eigenvalues close to the origin are compared to 
a complex generalization of the chiral Gaussian Symplectic Ensemble, 
confirming its predictions for weak and strong non-Hermiticity.
They differ from the QCD symmetry class with three colors
by a level repulsion from both the
real and imaginary axis.
\end{abstract}

\maketitle

\setcounter{footnote}{0}

\section{Introduction}

The comparison of small Dirac operator eigenvalues from lattice QCD 
to matrix models has been very successful at zero chemical 
potential~\cite{VW00}. 
At $\mu\neq0$ detailed comparisons to the 
microscopic Dirac operator origin spectrum have been few for two reasons. 
First, unquenched QCD suffers from a complex action problem and second, 
matrix model predictions where not available. 
Very recently substantial progress has been made for
complex chiral matrix models. 
Quenched~\cite{A02,SV04} and unquenched~\cite{James} predictions for QCD
have been achieved, as well as results
for adjoint gauge theories and staggered two color 
QCD~\cite{A04}. They have been successfully tested for 
quenched QCD~\cite{AW}, 
and our aim is to extend these results to other symmetry classes. 
The advantage of SU(2) is to avoid the complex determinant of the action
matrix~\cite{Hand99}. 
Already a few years ago, we computed the lowest eigenvalues of
the unquenched SU(2) Dirac operator at $\mu\neq0$~\cite{Bitt00a}, 
with low statistics 
though~\cite{Bitt01}. We redo the analysis as we can now compare to~\cite{A04}.
For simplicity we only present quenched results here.

\section{Matrix model predictions}

In Ref.~\cite{Step96} a complex chiral matrix model with 
the same global symmetries as QCD at $\mu\neq0$ was introduced. 
Its microscopic Dirac spectrum was obtained only very recently
\cite{SV04,James}, after previous asymptotic results in 
\cite{A02} for a related model. 
Although the presence of $\mu$ alters the chiral symmetry
breaking patterns for SU(2) and the adjoint representation, the 
symmetry analysis of Dirac matrix elements
being real, complex or quaternion remains valid~\cite{Hala97}.
Following the two-matrix model approach
\cite{James} new results for quaternionic matrices are now available
\cite{A04}. We only reproduce the quenched microscopic density 
here and refer to Ref.~\cite{A04} for more details.

In QCD at $\mu=0$, the lowest Dirac eigenvalues 
scale with the volume $1/V$  
to build up a finite condensate $\Sigma$ according 
to the Banks-Casher relation. 
At weak non-Hermiticity, a concept first introduced in~\cite{FKS97},
this scaling remains unchanged \cite{A02}. 
The support of the density remains quasi one-dimensional as we send $\mu\to0$,
keeping $\lim_{N\to\infty} 2N\mu^2 \equiv \alpha^2$ fixed. Here $N\sim2V$
is the size of the random matrix.
In contrast to that, at strong non-Hermiticity the eigenvalues fill a 
two-dimensional surface, and thus the scaling 
is modified to $\sim 1/\sqrt{V}$~\cite{A02}. 
The quenched microscopic density at weak non-Hermiticity 
in the sector of topological charge $\nu$ is reading~\cite{A04}
\begin{eqnarray}
\label{rhoweak}
\rho_{w}(\xi) &=& \frac{|\xi|^2(\xi^{\ast\,2}-\xi^2)}{2^4(\pi\alpha^2)^2}
K_{2\nu}\left(\frac{|\xi|^2}{\alpha^2}\right)
\mbox{e}^{\frac{\xi^2+\xi^{*\,2}}{2\alpha^2}}\\
&\times&\int_0^1 ds \int_0^1 dt\ t^{-\frac12}
\mbox{e}^{-s(1+t)\alpha^2}\nonumber\\
&\times&(J_{2\nu}(2\sqrt{st}\xi)J_{2\nu}(2\sqrt{s}\xi^\ast)
-(\xi\leftrightarrow \xi^\ast)).
\nonumber
\end{eqnarray}
Here, we have rescaled the complex eigenvalues according to $\xi\sim Vz$.

The limit of strong non-Hermiticity can be obtained by taking the limit
$\alpha \to\infty$ in Eq.~(\ref{rhoweak}). The resulting microscopic
density is symmetric with respect to the axes $\Re e(\xi)$ and $\Im m(\xi)$. 

The zero momentum sector of chiral perturbation theory with $\mu\neq0$ 
contains two parameters~\cite{SU2reference}, 
$\Sigma$ and the decay constant $f_\pi$, with the relative rescaling 
between mass and $\mu$ term depending on the ratio only. 
In the weak limit Eq. (\ref{rhoweak}) becomes a function of 
$\xi/\alpha$ only, by substituting $s\alpha^2=r$.
Note that the first integral then runs to $\alpha^2$.
However, at small $\alpha \gtrsim 0$ this change of variables breaks down.
The integral becomes approximately $\alpha$-independent while the prefactor
strongly varies with $\alpha$. Thus in this regime eigenvalues $z$ (or masses) 
and $\mu$ can be rescaled independently with {\it two} parameters.

\section{Lattice data at $\mu\neq0$}

Our computations with gauge group SU(2) on a $6^4$ lattice
at $\beta=4/g^2=1.3$ have $N_f=2$ flavors
of staggered fermions. For this system the fermion
determinant is real and lattice simulations of the full theory with
chemical potential are feasible. To compare with the quenched predictions 
from above we have chosen a large value for the quark mass $ma=20$ ($a$ 
being the lattice spacing in physical units) that 
effectively quenches the theory. We have checked this by comparing 
to simulations with $ma=2$, following exactly the 
same analytical curve as in Fig.~\ref{weakdata} below.
A  phase transition occurs 
at $\mu_c \approx m_{\pi}/2 \approx 0.3$ where the chiral condensate
goes to zero and a diquark condensate develops~\cite{Hand99}.

At strong gauge coupling and thus away from the continuum limit, 
staggered fermions have the disadvantage of shifting the (topological)
Dirac zero modes and mixing them with the nonzero modes. 
We have accounted for this by setting $\nu=0$ in 
Eq.~(\ref{rhoweak}) above. 
Only more sophisticated methods as improved actions~\cite{Follana}
can recover topology in the staggered approach.
We produced around 10000 configurations for two values of $\mu=0.001$
and $\mu=0.2$, corresponding to weakly and strongly
non-Hermitian lattice Dirac operators, respectively. 

Fig. \ref{weakdata} shows that 
the data agree very well to the functional form of Eq.~(\ref{rhoweak})
at weak non-Hermiticity. After normalizing histograms and formula to unity
we rescale the lattice eigenvalues 
$z$ by $\xi=\pi z/d$, with the spacing $d\propto 1/V$.
The weak non-Hermiticity parameter $\alpha^2$ has been rescaled 
with a second parameter $\sim f_\pi^2/d$ to match the data.
In contrast to QCD with three colors~\cite{A02,SV04,James}
the symplectic symmetry class shows
an additional suppression of eigenvalues along the real axis, resulting into
the double peak structure. 
The zero modes are driven away from the real axis for $\mu$ as small as
$\mu=0.000001$. This indicates that the limit $\mu \to 0$ is discontinuous.

\begin{figure}[-ht]
  \begin{center}
 \includegraphics[height=44.5mm]{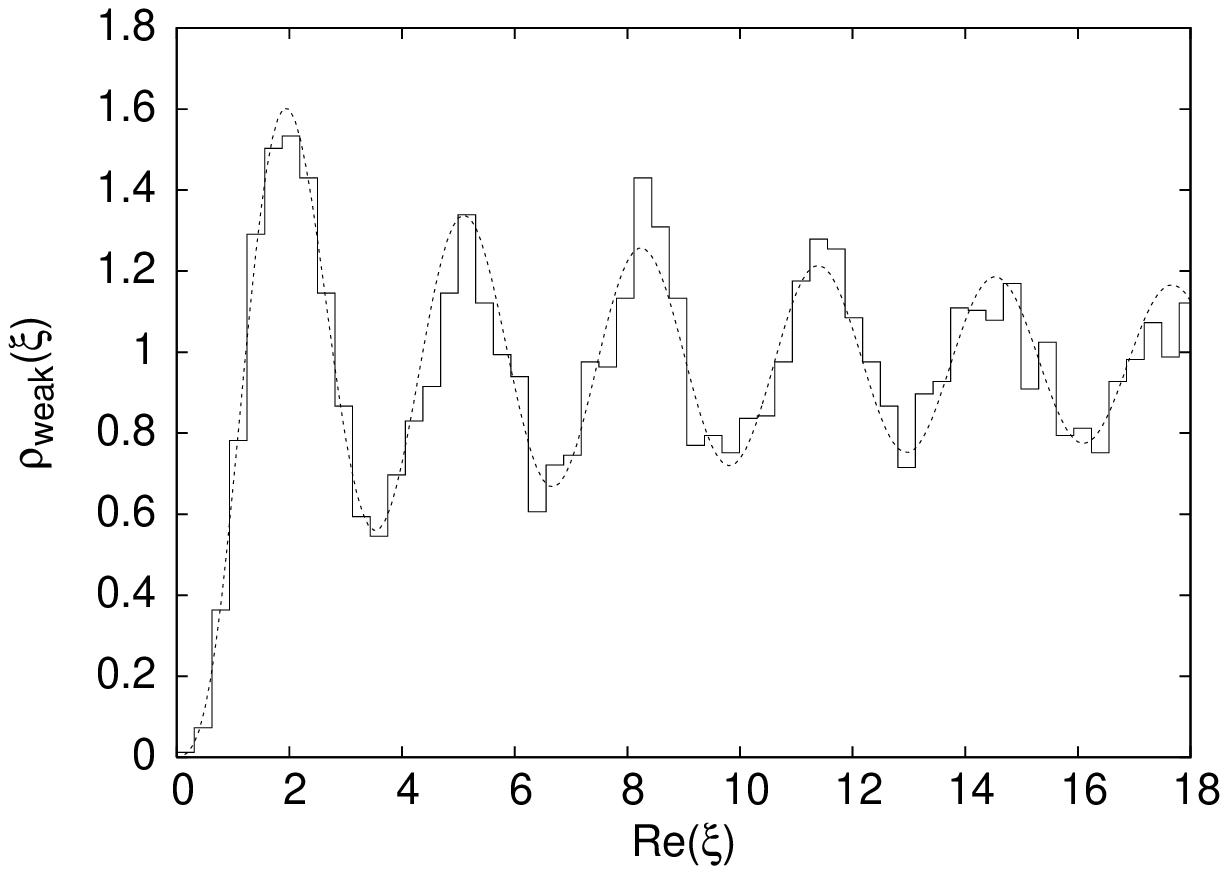}\\[0mm]
 \includegraphics[height=44.5mm]{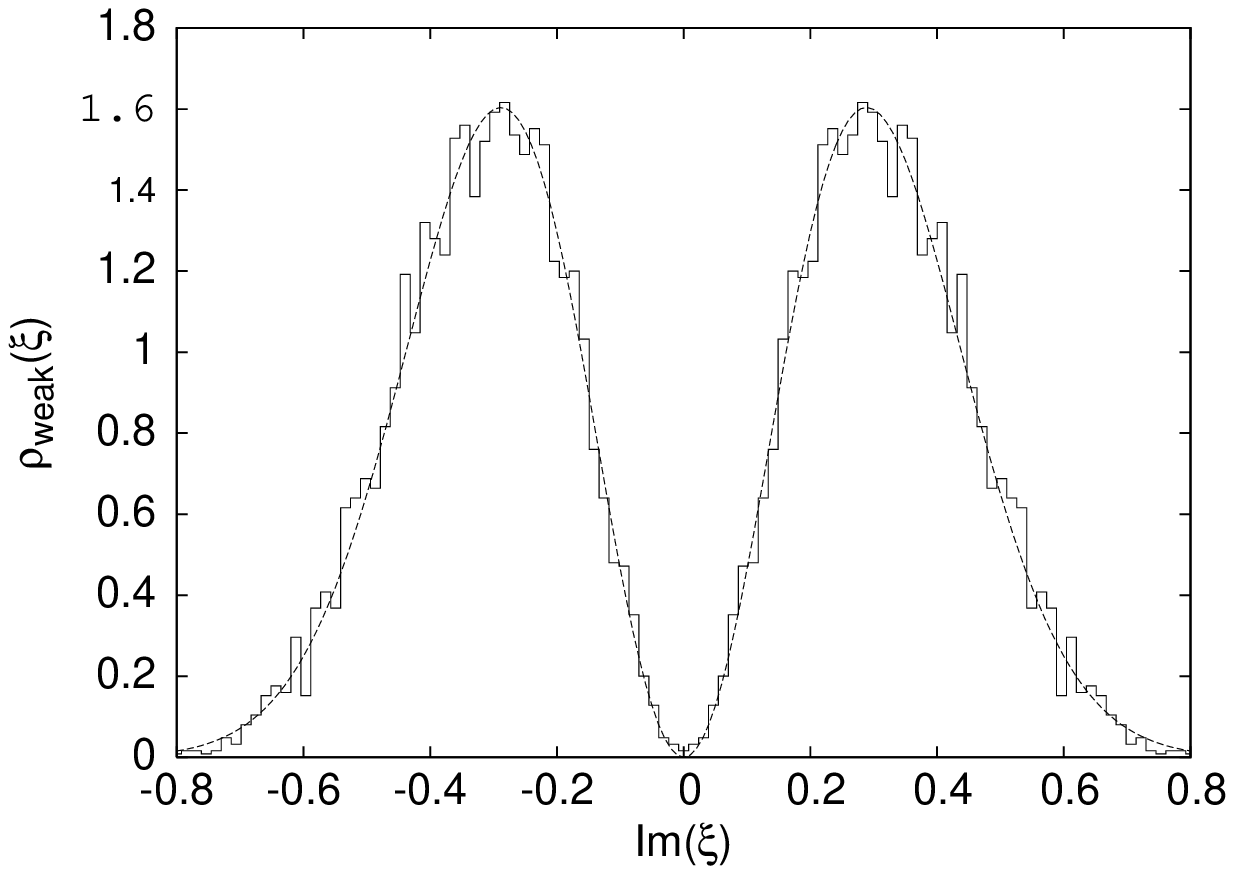}
  \end{center}
\vspace*{-11mm}
\caption{Data (histograms) for $V=6^4$ at $\mu=0.001$
vs.\ Eq.~(\ref{rhoweak}) for
weak non-Hermiticity ($\alpha\sim0.4$),
cut parallel to the real axis along maxima
(upper plot) and to the imaginary axis at the
first maximum (lower plot).}
\label{weakdata}
\vspace*{-6mm}
\end{figure}

\begin{figure}[-t]
  \begin{center}
    \includegraphics[height=44.5mm]{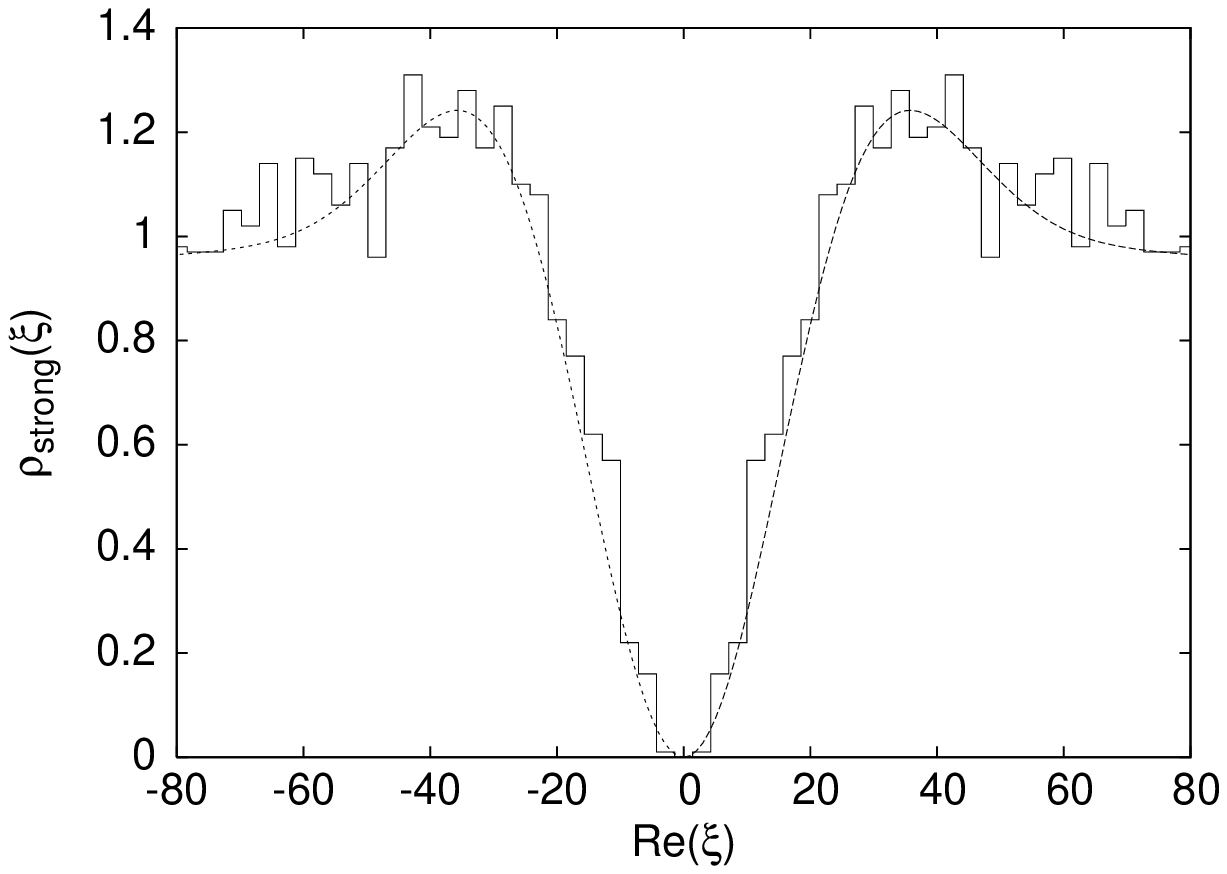}\\[0mm]
    \includegraphics[height=44.5mm]{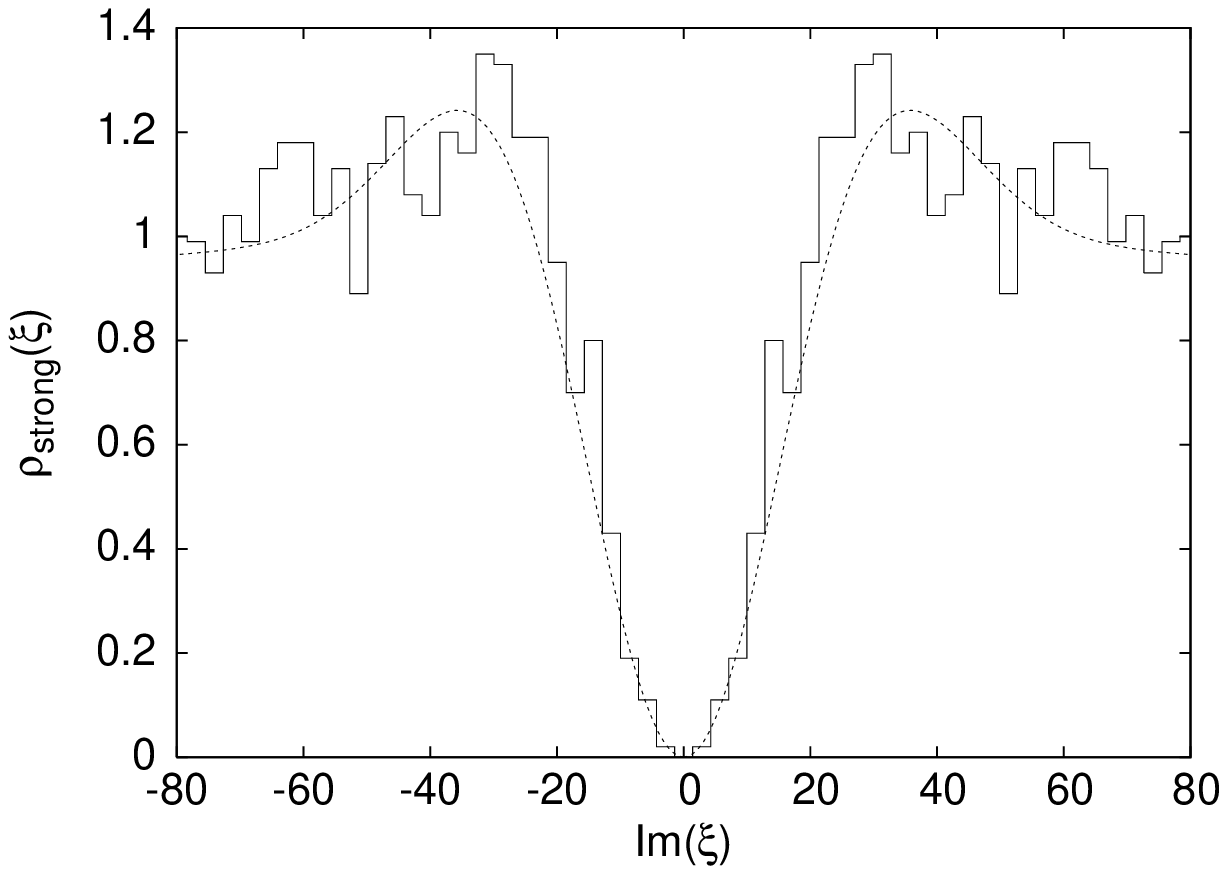}
  \end{center}
\vspace*{-11mm}
\caption{ Data (histograms) for $V=6^4$ at $\mu=0.2$
vs. Eq.~(\ref{rhoweak}) for strong non-Hermiticity ($\alpha\sim40$),
cut along the real (upper plot)
and imaginary (lower plot) axes.}
\label{strongdata}
\vspace*{-6mm}
\end{figure}

Fig.~\ref{strongdata} shows data for $ma=20$ at 
strong non-Hermiticity, with $\mu=0.2$ still below the phase
transition.
Here, the eigenvalues are rescaled with the square root 
of the volume, $\xi\sim z/\sqrt{d}$. Lacking an analytical 
prediction for the level spacing we have rescaled our eigenvalues 
by one parameter to fit Eq.~(\ref{rhoweak}).
The normalized data nicely confirm the prediction, showing the 
correct symmetry.

\section{Conclusions}

We have tested the analytical predictions 
from a complex chiral symplectic matrix model 
using effectively quenched two color QCD.
The agreement demonstrates that also at $\mu\neq0$ matrix models are an excellent tool 
to describe the different gauge theory symmetry classes. 
In this preliminary study we have established the correct 
qualitative behavior from quenched data. 
A detailed analysis including different
lattice volumes, scaling, and dynamical flavors
with small mass is in progress.\\

\noindent
{\bf Acknowledgments}: Part of this work was supported by a DFG Heisenberg
fellowship (G.A.). E.B.\ thanks the EU network HPRN-CT-1999-00161 EUROGRID -- 
``Geometry and Disorder: from membranes to quantum gravity'' for a postdoctoral grant.


\begin{thebibliography}{99}

\bibitem{VW00} J.J.M.~Verbaarschot and T.~Wettig,
{ Ann.\ Rev.\ Nucl.\ Part.\ Sci.} 50 (2000) 343.

\bibitem{A02} G. Akemann, Phys. Rev. Lett. 89 (2002) 072002;
  J. Phys. A 36 (2003) 3363. 

\bibitem{SV04} K. Splittorff and J.J.M. Verbaarschot,
Nucl. Phys. B 683 (2004) 467.

\bibitem{James} J.C. Osborn, hep-th/0403131.

\bibitem{A04} G. Akemann, preprint SPhT T04/069.

\bibitem{AW}  G. Akemann and T. Wettig, Phys. Rev. Lett. 92 (2004) 102002.

\bibitem{Hand99} S. Hands, J.B. Kogut, M.-P. Lombardo, and
  S.E. Morrison, Nucl. Phys. B 558 (1999) 327.

\bibitem{Bitt00a} E. Bittner, M.-P. Lombardo, H. Markum, and
R. Pullirsch,
Nucl. Phys. B (Proc. Suppl.) 94 (2001) 445.

\bibitem{Bitt01}
E. Bittner, M.-P. Lombardo, H. Markum, and R. Pullirsch,
Nucl. Phys. B (Proc.~Suppl.) 106 (2002) 468;
Spec. Issue Nucl. Phys. A 702 (2002) p12.

\bibitem{Step96} M.A. Stephanov, Phys. Rev. Lett. 76 (1996) 4472.

\bibitem{Hala97} M.A. Halasz, J.C. Osborn, and J.J.M. Verbaarschot,
                 Phys. Rev. D 56 (1997) 7059.

\bibitem{FKS97} Y.V. Fyodorov, B.A. Khoruzhenko, and H.-J. Sommers,
  Phys. Lett. A 226 (1997) 46;
  Phys. Rev. Lett. 79 (1997) 557.

\bibitem{SU2reference}
J.B. Kogut, M.A. Stephanov, D. Toublan, J.J.M. Verbaarschot, and A. Zhitnitsky,
Nucl. Phys. B 582 (2000) 477.

\bibitem{Follana}
E. Follana, A. Hart, and C.T.H. Davies, hep-lat/0406010.

\end{thebibliography}
\end{document}